\documentclass{phyeauth}

\usepackage{epsfig}
\usepackage{amsmath}
\usepackage{amssymb}

\begin{document}

\newcommand{\ugl}{\ensuremath{U_\text{gL}}}
\newcommand{\ugr}{\ensuremath{U_\text{gR}}}
\newcommand{\ugc}{\ensuremath{U_\text{gC}}}
\newcommand{\usd}{\ensuremath{U_\text{SD}}}
\newcommand{\usddc}{\ensuremath{U_\text{SD,dc}}}
\newcommand{\un}[1]{\,\text{#1}}
\newcommand{\bper}{\ensuremath{B_\perp}}
\newcommand{\mus}{\ensuremath{\mu_\text{S}}}
\newcommand{\mud}{\ensuremath{\mu_\text{D}}}
\newcommand{\mul}{\ensuremath{\mu_\text{L}}}
\newcommand{\mur}{\ensuremath{\mu_\text{R}}}
\newcommand{\Ia}{\ensuremath{\text{I}^\ast}}

\begin{frontmatter}

\title{Molecular states in a one--electron double quantum dot}

\author[lmu]{A.K.~H\"uttel}\ead{mail@akhuettel.de},
\author[lmu]{S. Ludwig},
\author[lmu]{H. Lorenz},
\author[fkf]{K.~Eberl\thanksref{lumix}},
\author[lmu]{J.P.~Kotthaus}

\address[lmu]{Center for NanoScience and Department Physik,
Ludwig--Maximilians--Universit\"at, Geschwister--Scholl--Platz 1,
80539~M\"unchen, Germany}

\address[fkf]{Max-Planck-Institut f\"ur Festk\"orperforschung,
Heisenbergstra{\ss}e 1, 70569 Stuttgart, Germany}

\thanks[lumix]{Present address: Lumics GmbH, Carl--Scheele--Stra{\ss}e
  16, 12435 Berlin, Germany.}

\begin{abstract} 
The transport spectrum of a strongly
tunnel-coupled one-electron double quantum dot electrostatically
defined in a GaAs/AlGaAs heterostructure is studied. At finite
source-drain-voltage we demonstrate the unambiguous identification of the
symmetric ground state and the antisymmetric excited state of the double well
potential by means of differential conductance measurements. A sizable
magnetic field, perpendicular to the two-dimensional electron gas, reduces the
extent of the electronic wave-function and thereby decreases the tunnel
coupling. A perpendicular magnetic field also modulates the orbital
excitation energies in each individual dot. By additionally tuning the
asymmetry of the double well potential we can align the chemical potentials of
an excited state of one of the quantum dots and the ground state of the other
quantum dot. This results in a second anticrossing with a much larger tunnel
splitting than the anticrossing involving the two electronic ground states. 
\end{abstract}

\end{frontmatter}

Electrostatically defined semiconductor double quantum dots, where
electrons are confined in a double potential well, have recently attracted
considerable attention \cite{rmp-wiel:1}. The interest in these artificial
molecules is largely due to the proposed use of quantum dots as spin or
charge qubits, the building blocks of the hypothetical quantum computer
\cite{pra-loss:120,jjap-vanderwiel:2100}.  Recent works have shown
spectacular advancements in reducing the number of electrons trapped in a
double quantum dot (DQD) down to $N=1$
\cite{prb-elzerman:161308,prl-petta:186802,anticrossing,cm-pioro-ladriere:0504009}.
Here we study the transport spectrum of a strongly tunnel-coupled DQD
with $N\le 1$ at finite source-drain voltage \usd. We observe
molecule-like hybridization not just between the ground states of both
quantum dots, but also at finite potential asymmetry between the ground
state of one quantum dot and an excited state of the other dot.

The measurements have been performed on an epitaxially grown
AlGaAs/GaAs heterostructure forming a two-dimensional electron system
(2DES) $120\,\text{nm}$ below the crystal surface. The electron sheet
density in the 2DES is $n_\text{s} = 1.8\times 10^{15} \text{m}^{-2}$,
the electron mobility $\mu = 75 \,\text{m}^2/\text{Vs}$. We estimate the
2DES electron temperature to be of the order $T_\text{2DES} \simeq
100\,\text{mK}$.
\begin{figure}[tb]\begin{center}
\epsfig{file=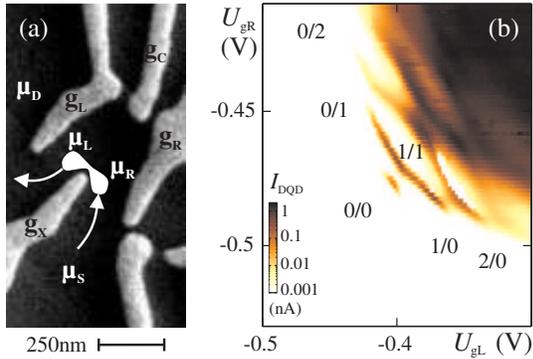, width=7cm}
\end{center}
\caption{ (a) SEM micrograph of the gate electrode geometry used to
  define a DQD. The approximate position of the DQD and the current
  path is indicated in white. (b) Current through the DQD as function
  of the side gate voltages \ugl\ and \ugr\ ($\usddc =
  50\,\mu\text{V}$, logarithmic color scale).}
\label{fig1}
\end{figure}
Fig.~\ref{fig1}(a) displays an electromicrograph of the gate structure
on the crystal surface used to electrostatically define a DQD.
The layout is based on the triangular geometry for single
quantum dots at very low electron numbers introduced by Ciorga {\it et
al.} \cite{prb-ciorga:16315}. By tuning the voltages on center gates
$\text{g}_\text{C}$ and $\text{g}_\text{X}$ to increasingly negative
values, we deform the trapping potential in order to create two
potential minima shaping a DQD. The approximate
geometry of this DQD is indicated in Fig.~\ref{fig1}(a)
by a white peanut-like shape. Its two quantum dots are strongly
tunnel-coupled to each other with a tunnel splitting of typically
$0.03\un{meV} \lesssim 2t_0 \lesssim 0.3\un{meV}$ \cite{anticrossing}.

Fig.~\ref{fig1}(b) displays the dc current through the DQD
in linear response ($\usddc = 50\,\mu\text{V}$) as function of the
side gate voltages \ugl\ and \ugr. The hexagons of Coulomb blockade
typical for transport through a DQD can be recognized
\cite{rmp-wiel:1}. Charge sensing measurements using a nearby quantum
point contact provide proof that in the area marked 0/0 the DQD
is entirely depleted of conduction band electrons \cite{anticrossing}.
The subsequent regions of increasing charge in each dot are marked by
pairs of numbers $N_L/N_R$, where $N_L$ ($N_R$) is the absolute number of
electrons trapped in the left (right) quantum dot. For a weakly tunnel
coupled DQD such a stability diagram shows current only at
the sharp hexagon corners where three different charge configurations are
energetically possible (triple points) \cite{rmp-wiel:1}. In contrast, in
Fig.~\ref{fig1}(b) we observe single electron tunnelling (SET) even along
the hexagon lines connecting triple points. These resemble not sharp but
rounded hexagon corners. This indicates delocalized electronic
states due to strong tunnel coupling between the two dots.

\begin{figure}[tb]
\begin{center} 
\epsfig{file=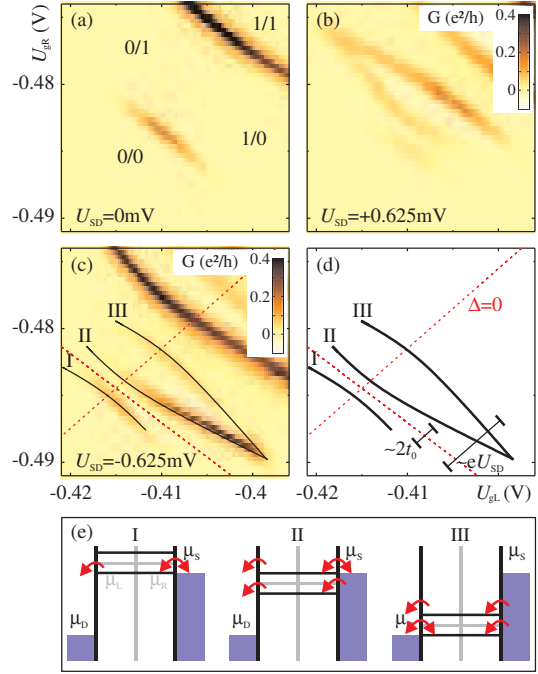, width=7cm}
\end{center} 
\caption{ Expansion of the first triple point of the stability diagram
  at finite source-drain voltage. (a), (b), (c): Differential
  conductance at $\usddc = 0$ and  $\usddc=\pm 0.625\un{mV}$, 
  with model lines added in (c). (d)
  Corresponding model expectations (see text, $\usddc=-0.625\un{mV}$,
  $2t_0 = 0.2\un{meV}$). (e) Level alignment schemes, showing the
  chemical potentials of source \mus, drain \mud, and the energies of
  the molecular states. The three graphs correspond to the
  intersections of lines I, II, and III in (d) with the $\Delta=0$
  line of symmetric double well potential.}
\label{fig2}
\end{figure}

In this article, we focus on transport that takes place through
one-electron quantum states, i.e. the region of the stability diagram
where the charge configurations 0/0, 1/0, and 0/1 are accessible.
Fig.~\ref{fig2}
compares the differential conductance of this region of the stability
diagram for zero source-drain voltage $\usddc=0$ in (a) with the same
region for $\usddc=\pm 0.625\un{mV}$ in (b) and (c). In linear
response (Fig.  \ref{fig2}(a)) the conductance exhibits the same
behaviour as the current shown in Fig. \ref{fig1}(b), i.e. the lines
of high current match the local differential conductance maxima (dark
lines).

In the case of weak interdot coupling, the triple points of the stability
diagram expand at finite source-drain voltage to triangular regions of
finite current \cite{rmp-wiel:1,cm-johnson}, or a triangle in differential
conductance. Here, i.e. for strong tunnel coupling, a more complex
structure of three curved lines is observed.  The three lines, marked for
$\usddc=-0.625\un{mV}$ in Figs. \ref{fig2}(c) and (d) with I, II, and III,
correspond to steps in the SET current and indicate that a delocalized
quantum level of the DQD is aligned with the chemical
potentials of either the source or the drain lead. The detailed situations
leading to maximum differential conductance are schematically drawn in
Fig. \ref{fig2}(e): Along line I, tunneling through the symmetric ground
state of the double well potential becomes accessible, as its energy
matches the chemical potential in the source lead \mus\ (left plot). Line
II is caused by an increase in current as the antisymmetric first excited
state of the double well potential enters the transport window, 
providing a second transport channel (middle plot). Along line III the ground state drops
below the drain chemical potential \mud\ (right plot). For even higher gate voltages,
the ground state is permanently occupied, such that Coulomb blockade
prohibits SET. Since the same quantum state is involved in both cases,
lines I and III are parallel to each other.

Obviously, the distance between lines I and II corresponds to the
excitation energy $2\sqrt{t_0^2 + \Delta^2}$, where $2\Delta$ is the
potential asymmetry in the DQD with $2\Delta = ( \mur -
\mul )$. In comparison, the distance between lines I and III corresponds
to the difference in chemical potential between source and drain contact
$e \usd$ and provides a known energy scale. Lines I and II depict the
anticrossing due to the hybridization of the two orbital ground states of
both quantum dots. The solid model lines in Figs. \ref{fig2}(c)--(d)
resemble the energy splitting $2\sqrt{t_0^2 + \Delta^2}$ and are obtained
using a tunnel splitting of $2t_0=0.2\un{meV}$. The model lines have been
transformed from the energy to the gate voltage scale by taking into
account the geometrical capacitances between gates and quantum dots. The
latter were obtained from the slopes of lines of maximum differential
conductance similar as in Ref$.$ \cite{anticrossing}. Note, that this is a
linear transformation, thus, allowing the determination of $2t_0$ simply by
comparison of the smallest distances between lines I and II versus lines I
and III.

\begin{figure}[tb]
\begin{center} 
\epsfig{file=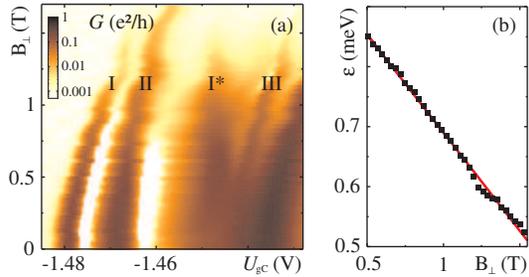, width=7cm}
\end{center} 
\caption{(a) Differential conductance $G$ as function of center gate
  voltage \ugc\ and magnetic field \bper, for slightly asymmetric
  potential in the DQD and $\usddc = -1\un{mV}$.  Lines I, II, and III
  are marked as in Fig. \ref{fig2}. A higher excited quantum state is
  visible through line \Ia. (b) Excitation energy of this state as
  function of \bper.}
\label{fig3} 
\end{figure}

At large enough source-drain voltage (large transport window) an
additional excited orbital state is observed that decreases in energy with
increasing magnetic field \bper\ perpendicular to the 2DES. 
This is demonstrated in Fig. \ref{fig3}(a),
where the differential conductance is plotted as a function of center gate
voltage \ugc\ (see Fig. \ref{fig1}(a)) and \bper\ for a rather large
$\usddc=-1.0\un{mV}$. Gate $\text{g}_\text{C}$ couples approximately
symmetrical to both quantum dots. The side gate voltages \ugl\ and
\ugr\ are adjusted such that $|\Delta|\lesssim 0.1\un{meV}$ is provided
throughout Fig. \ref{fig3}(a). Lines I, II, and III can be
identified with the lines marked respectively in Fig. \ref{fig2}. Between
lines II and III an additional line of enhanced differential conductance,
marked \Ia, becomes visible. It represents a transport channel
corresponding to an additional excited orbital state of one of
the two quantum dots. The broad
dark line at the right edge of the plot marks the onset of
tunneling through two-electron states with $1\le N\le 2$.

The excitation energy $\epsilon$ of the excited state causing line \Ia\
corresponds to the distance between the conductance maxima of lines I and
\Ia. This energy $\epsilon$ is plotted in Fig. \ref{fig3}(b) as function
of the magnetic field for $0.5\un{T} \le \bper \le 1.5\un{T}$. In this
field range line \Ia\ yields an isolated conductance maximum. The solid
line depicts $\epsilon = 1.03\un{meV}- 0.34
\,\frac{\text{meV}}{\text{T}} \bper$ suggesting a linear dependence of
$\epsilon$ on the magnetic field \cite{kink}.

\begin{figure}[tb]
\begin{center} 
\epsfig{file=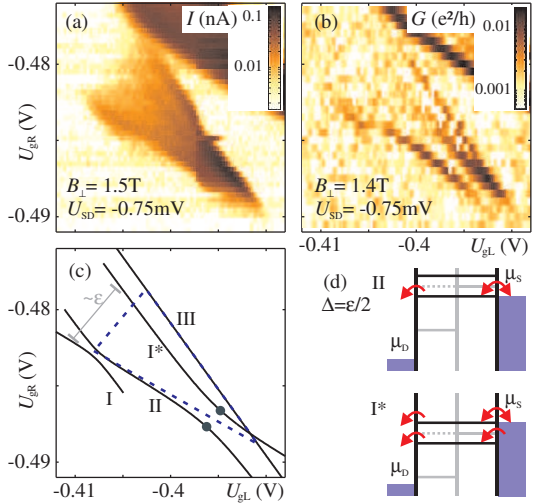, width=7cm} 
\end{center}
\caption{ The first triple point of the charging diagram at $\bper
  \simeq 1.5\un{T}$ and $\usddc=-0.75\un{mV}$. (a) dc current, (b)
  differential conductance (logarithmic color
  scale), (c), (d) Model lines and level alignment
  schemes for an additional level anticrossing (see text for
  details).}
\label{fig4} 
\end{figure}

Fig. \ref{fig4} displays the transport spectrum at the first triple point for
$\usddc=-0.75$ mV and $\bper \simeq 1.5\un{T}$.
At such a
high magnetic field the tunnel splitting caused by the hybridization of both
quantum dot ground states is decreased to almost zero because of the increased
localization of the orbital wave functions in a perpendicular magnetic field
\cite{anticrossing} (comp. lines I and II in Fig. \ref{fig4}(c)). Therefore, the
region of high current in Fig. \ref{fig4}(a) marking the first triple point of
the stability diagram resembles a triangle as expected
for electronic states almost localized within the two quantum dots. However, the
tip of the triangle seems distorted and shows increased current. The reason for
this is revealed by the corresponding differential conductance measurement shown
in Fig. \ref{fig4}(b). It depicts an anticrossing of lines II and \Ia\ near the
tip of the triangle.

A model describing these observations is plotted in Fig. \ref{fig4}(c). The
model lines assume a ground state -- ground state tunnel coupling of $2t_0
\simeq 0.064\un{meV}$. An excited orbital state of the left dot (line \Ia) has
an excitation energy $\epsilon=0.55\un{meV}$.
It hybridizes with the ground state of the right quantum dot for a potential
asymmetry $2\Delta = \epsilon$ that makes these two states energetically
degenerate. Both lines \Ia\ in Fig. \ref{fig3} and Fig. \ref{fig4}
correspond consistently to the same excited state in the left dot.
For a tunnel splitting of $2t_0^\ast=0.2\un{meV}$, describing the
second anticrossing, the model lines of Fig. \ref{fig4}(c) show good agreement with the observed
differential conductance maxima. The delocalized states generated by such a
hybridization also provide a good explanation for the enhancement of SET as
observed in Fig. \ref{fig4}(a). Note, that the
tunnel coupling $2t_0^\ast \gg 2t_0$ is sizable even for the large magnetic
field of $\bper \simeq 1.5\un{T}$. This can be explained in terms
of a smaller effective tunnel barrier between the quantum dots for excited orbital
states. Possible causes include the higher energy of the excited orbital
state or, alternatively, a different orbital symmetry of
the excited state, allowing stronger coupling between the quantum dots.

In conclusion, we directly observe anticrossings of molecular states, as a
consequence of the quantum mechanical tunnel coupling of one-electron orbital
states in two adjacant quantum dots. A conductance measurement at finite source
drain voltage reveals the molecular symmetric and antisymmetric states,
resulting from the tunnel coupled orbital ground states in both dots, as
distinct lines in the stability diagram. A large perpendicular magnetic field
quenches this anticrossing. Strikingly, at a large
perpendicular magnetic field and finite potential asymmetry we find a second
sizable anticrossing between the ground state of one dot and an excited orbital
state of the other dot.

We thank R.\ H.\ Blick, U.\ Hartmann, and F.\ Wilhelm for valuable
discussions, and S.\ Manus for expert technical help, as well as the
Deut\-sche For\-schungs\-ge\-mein\-schaft for financial
support.  A.\ K.\ H.\ thanks the German Nat. Academic Foundation
for support.

\bibliographystyle{elsart-num}
\bibliography{paper}

\begin{thebibliography}{10}
\bibitem{kink}
The tiny kink in the data course at $\bper \simeq 1.2\un{T}$ may be related to
  the fact that here the two-dimensional electron gas reaches the filling
  factor $\nu=6$, leading to shifts in all line positions.
\end{thebibliography}

\end{document}